\documentstyle{mn} 
\input epsf

\title[Wide-angle tailed radio sources]
{The origin of wide-angle tailed radio galaxies}
\author[I.~Sakelliou and M.R. Merrifield]
          {Irini~Sakelliou$^{1}$, Michael R.~Merrifield$^{2}$\\ 
$^{1}$ Mullard Space Science Laboratory, University College
London, Holmbury St Mary, Dorking, Surrey RH5 6NT, UK\\
$^{2}$ School of Physics and Astronomy, University of
Nottingham, University Park, Nottingham, NG7 2RD\\
}

\begin{document}
\maketitle
 
\begin{abstract} 

To investigate the origins of wide-angle tailed radio sources (WATs),
we have compiled a sample of these systems in Abell clusters for which
X-ray data exist.  Contrary to conventional wisdom, the WATs are found
to be significantly displaced from the X-ray centroids of their host
clusters.  The bends in the WATs' radio jets are found to be oriented
preferentially such that they point directly away from or toward the
cluster centre, with more of the former than the latter.  If this
morphology is attributed to ram pressure, then the WATs are on
primarily radial orbits, with more approaching the X-ray centroid than
receding.  There is also some evidence that the in-coming WATs are
on average further from the X-ray centroid than the out-going ones.
All of these observations strongly support a scenario in which WATs
are created in cluster mergers.

\end{abstract}
 

\section{Introduction}

Wide angle tail radio galaxies (WATs) form a class of radio galaxies,
usually found in clusters, whose radio-emitting jets have been bent
into a wide `C' shape.  This structure gives the immediate impression
that the jets are being swept back by the dynamic pressure resulting
from the motion of the associated galaxy through the surrounding
intracluster medium (ICM).  This `ram pressure' model was first
developed by Begelman, Rees \& Blandford (1979), and studied in more
detail by Vall\'{e}e, Bridle \& Wilson (1981) and Baan \& McKee
(1985).

Unfortunately, there is a piece of evidence that seems to contradict
this intuitively-appealing model.  WATs are usually associated with
the brightest clusters ellipticals (D or cD galaxies), and these
galaxies are generally found at rest, close to the centres of clusters
(Quintana \& Lawrie 1982, Bird 1994, Pinkney 1995).  From a
theoretical point of view, this finding can be understood since models
of cluster formation imply that large galaxies form close to their
cluster centres (Bode et al.\ 1994; Garijo, Athanassoula \&
Garc\'{\i}a-G\'{o}mez 1997).  Even if a massive galaxy were initially
placed on a high-velocity orbit that carried it far out in its
cluster, dynamical friction would rapidly drag it down to rest at the
centre of the system (Ostriker \& Tremaine 1975).

Since it seems that the D/cD galaxies that host WATs should lie at
rest in the centres of their clusters, they should not possess the
motion required to produce the observed bends in their radio jets by
ram pressure.  It has therefore been thought necessary to invoke
alternative mechanisms to explain the observed bends in WATs' jets.
One candidate for this mechanism is an electromagnetic force arising
from the interaction between a jet that carries a net electrical
current and the magnetic field in the ICM (Eilek et al.\ 1984).  Given
our poor understanding of currents in jets and magnetic fields in
clusters, this model has not been extensively explored.  One problem
with it is that it requires a highly and favourably ordered magnetic
field in order to produce the symmetric shape of WATs.  Alternatively,
jets could be deflected by collisions with dense clouds in the ICM.
Although this process may be at work in some radio galaxies whose jets
are deflected and disrupt abruptly (Burns et al.\ 1986), again it has
difficulty reproducing the large-scale symmetric structure of WATs.
Thus, neither of the suggested alternative jet-bending mechanisms are
entirely satisfactory.

A possible solution to this dilemma has come from the realization that
clusters are dynamically young, and merge frequently.  Theoretical and
observational studies have forced us to discard the idealized picture
of a spherical relaxed cluster that is isolated and does not interact
with its surroundings.  Instead, structure in the Universe is now
viewed as evolving hierarchically, with large feature such as clusters
forming through the repeated mergers of smaller groups
(e.g. Evrard 1990; Jing et al.\ 1995; Frenk et al.\ 1996).

It has therefore been suggested that the galaxy motions required to
bend WATs by ram pressure are a by-product of collisions between
clusters (Pinkney, Burns \& Hill 1994; G\'{o}mez et al.\ 1997a, b;
Loken et al.\ 1995).  Consider a radio galaxy located at the centre of
a cluster.  If this cluster collides with a second
comparable system, then the collisional nature of the ICM means that
the kinetic energy of the gas will rapidly dissipate, and the two
separate gaseous components will merge into a single structure.  The
radio galaxy, on the other hand, is an essentially collisionless
system that will not be decelerated at the same rate as the
surrounding ICM, and so it will be kicked into motion as efficiently
as a passenger in a car accident who is not wearing a seat belt.  The
motion of the galaxy relative to the ICM will then generate the ram
pressure needed to bend the radio jets.  Some recent support for this
idea has come from the work of Novikov et al.\ (1999), who found that
the jets of WAT radio galaxies tend to be aligned with the long axis
of any surrounding supercluster; it is along this axis that one would
expect cluster mergers to occur preferentially (Colberg et al.\ 1999).

In this paper, we make direct observational tests of the merger
theory by investigating the properties of a sample of WAT sources.
The simplest prediction of this scenario is that, unlike most
D/cD galaxies, the hosts of WATs will not generally lie at the centre
of their clusters.  This prediction can most readily be tested by
comparing the location of the WAT to that of the centroid of the
cluster's X-ray emission, which should lie close to the global minimum
of the cluster's potential.\  In a major merger, the X-ray emission is
likely to be significantly disturbed, with shock heating at the
collision interface, and large-scale flows set up in the combined ICM,
so some care must be taken in equating the X-ray centroid with the
mass centre of the cluster.  However, the global distribution of X-ray
emitting gas will still reflect the morphology of the merging system,
so a centroid measured in a manner that is not heavily weighted toward
localized features should still provide a reasonable estimate of the
cluster centre.

Once the centroid has been determined, we can also determine the
direction in which the radio jets are bent relative to the cluster
centre.  If ram pressure is responsible for this morphology, then we
can use this information to study the orbits followed by the WATs, to
see if they are consistent with the cluster merger model.

The remainder of this paper, describing these tests, is laid out as
follows.  Section~2 presents the process by which radio galaxies were
selected to represent the WAT class.  Section~3 describes the X-ray
observations of the clusters containing these systems, and Section~4
presents the analysis that we have applied to them in order to measure
the centroids and spatial extents of their ICMs.  In Section~5, we
discuss the spatial distribution of the WATs relative to their host
clusters as inferred from the X-ray data, and we quantify
the orbits that the WATs follow as derived from the directions in
which the radio jets are bent.  Section~6 summarizes the findings, and
discusses their implications for the merger theory.

\section{Sample selection}

\begin{table*}
\begin{center}
 \caption{WATs in Abell clusters}
 \label{tab:WATs}
\vspace{0.5cm}
\small
 \begin{tabular}{lllrlrlc}   \hline \hline
\multicolumn{1}{c}{(1)} & 
\multicolumn{1}{c}{(2)} & 
\multicolumn{1}{c}{(3)} & 
\multicolumn{1}{c}{(4)} & 
\multicolumn{1}{c}{(5)} & 
\multicolumn{1}{c}{(6)} & 
\multicolumn{1}{c}{(7)} \\
       &
\multicolumn{2}{c}{\underline{Optical Position(1950)}}  & 
 &
&
\multicolumn{2}{c}{\underline{{\it ROSAT} OBS.}} &\\
\multicolumn{1}{c}{Source} & 
\multicolumn{1}{c}{R.A.} & 
\multicolumn{1}{c}{DEC} & 
\multicolumn{1}{c}{Abell} & 
\multicolumn{1}{c}{z} & 
\multicolumn{1}{c}{Sequence} & 
\multicolumn{1}{c}{Exposure}\\ 
\multicolumn{1}{c}{name} & 
\multicolumn{1}{c}{$^{\rm h} \; ^{\rm m}\; ^{\rm s}$}   & 
\multicolumn{1}{c}{$\degr \; \arcmin \; \arcsec$} & 
\multicolumn{1}{c}{cluster} & 
\multicolumn{1}{c}{}  & 
\multicolumn{1}{c}{number} & 
\multicolumn{1}{c}{(sec)} \\
\hline
 
0035+180 & 00 35 17.19  & +18 04 23.9 & A69 & 0.1448  & -- & \\

0110+152 & 01 10 20.45 & +15 13 35.2 & A160 & 0.0444 & rh800953 &
16501 & \\

0123--016B, 3C~40 & 01 23 27.55 & --01 36 18.9 & A194 & 0.0180 &
rp800316 & 24482 & \\

0141+061 & 01 41 19.20 & +06 09 34.0 & A245 & 0.0788 & -- & &\\

0146+138 & 01 46 30.66 & +13 48 08.2 & A257 & 0.0706 & -- & \\

0255+058A, 3C~75 & 02 55 02.99 & +05 49 37.0 & A400 & 0.0238 &
rp800226 & 23615 & \\
0255+058B & 02 55 03.08 & +05 49 20.9 & \\

0327+246B & 03 27 32.25 & +24 37 36.0 & A439 & 0.1063 & -- & & \\

0647+693, 4C~69.08 & 06 47 54.58 & +69 23 31.5 & A562 & 0.1100 &
rp800500 & 5865 & \\

0658+330 & 06 58 52.48 & +33 01 47.2 & A567 & 0.1270 & -- & & \\

0705+486 & 07 05 21.39 & +48 41 47.2 & A569 & 0.0195 & wp800575 & 4181 & \\

0803--008, 3C~193 & 08 03 05.14 & --00 49 43.1 & A623 & 0.0871 &
rp800506 & 4447 &  \\

0836+290 & 08 36 13.50 & +29 01 12.9 & A690 & 0.0788 & -- & & \\

0838+32, 4C~32.26 & 08 38 06.74 & +32 35 42.2 & A695 & 0.0694 & -- & \\

0908--103 & 09 08 32.61 & --10 21 33.7 & A761 & 0.0921 & -- & \\

0909+162 & 09 09 48.50 & +16 12 23.0 & A763 & 0.0851 & -- & \\

1011+500 & 10 11 26.68 & +50 00 27.6 & A950 & 0.2081 & rp700220 & 5151\\

1025+040 & 10 25 47.94 & +04 00 52.1 & A1024 & 0.0733 & -- & \\

1108+410A & 11 08 54.20 & +41 03 25.7 & A1190 & 0.0794 & -- & \\

1131+493, IC~708 & 11 31 16.25 & +49 20 19.8 & A1314 & 0.0338 &
wp800392 & 2940 & \\

1159+583, 4C~58.23 & 11 59 30.41 & +58 18 51.3 & A1446 & 0.1035 &
rp800501 & 7094 &\\

1200+519 & 12 00 34.13 & +51 57 12.4 & A1452 & 0.0630 & -- & \\

1221+615 & 12 21 07.38 & +61 31 29.6 & A1529 & 0.2324 & -- & \\

1225+636 & 12 25 33.20 & +63 39 37.8 & A1544 & 0.1459 & -- & &\\

1227+119 & 12 27 20.34 & +11 57 13.1 & A1552 & 0.0843 & wp800577 & 3398 & \\

1231+674, 4C~67.21 & 12 31 03.88 & +67 24 17.2 & A1559 & 0.1071 & -- & & \\

1233+168, 4C~16.33 & 12 33 55.19 & +16 48 47.6 & A1569 & 0.0784 &
rp800504 & 3687 & \\

1243+26(7) & 12 43 54.69 & +26 43 39.3 & A1609 & 0.0891 & -- & & \\

1300+32(1) & 13 00 54.54 & +32 06 08.1 & A1667 & 0.1648 & -- & &   \\

1306+107A, 4C~10.35 & 13 06 34.40 & +10 45 32.8 & A1684 & 0.0864 & -- & & \\

1320+584 & 13 20 58.62 & +58 25 41.3 & A1731 & 0.1932 & -- & & \\

1333+412, 4C~41.26 & 13 33 09.52 & +41 15 24.1 & A1763 & 0.2074 &
rp800252 & 15659 & \\

1415+084 & 14 15 02.92 & +08 26 19.8 & A1890 & 0.0570 & rh800649 & 13836 & \\

1433+553 & 14 33 54.92 & +55 20 53.2 & A1940 & 0.1396 & rp800502 & 3433 &  \\

1445+149 & 14 45 40.62 & +14 59 19.5 & A1971 & 0.2084 & -- & & \\

1636+379 & 16 36 15.71 & +37 58 53.7 & A2214 & 0.1610 & rp800503 & 5209 &  \\

1638+538, 4C~53.37 & 16 38 24.50 & +53 52 30.8 & A2220 & 0.1106 &
rp201446 & 24402 & \\

1820+689 & 18 20 01.32 & +68 55 24.0 & A2304 & 0.0880 & rp800498 & 5412 &  \\

1826+747 & 18 26 23.40 & +74 42 05.8 & A2306 & 0.1271 & rp800505 & 4470 & \\

2236-176 & 22 36 30.07 & --17 36 04.8 & A2462 & 0.0698 & rp800495 & 4289 &  \\

2330+091 & 23 30 58.81 & +09 08 58.7 & A2617 & 0.1623 & -- & & \\

2335+267, 3C~465 & 23 35 58.93 & +26 45 16.2 & A2634 & 0.0321 &
rp800014 & 20400 
 &  \\

2336+212 & 23 36 11.04 & +21 13 26.3 & A2637 & 0.0707 & rp000111 & 2580 &   \\
\hline
 
 \end{tabular}
\end{center}
\end{table*}

Over the last ten years, Abell clusters of galaxies have been
extensively mapped at the radio frequency of 1.4~GHz by Zhao, Burns \&
Owen (1989), Owen, White \& Burns (1992), and Owen, White \& Ge
(1993).  Maps of the radio sources found in these clusters, along with
the optical identifications of the galaxies that host them, are given
by the previous works and by Ledlow \& Owen (1995a) and Owen \& Ledlow
(1997).  This sample of radio galaxies in Abell clusters is complete
for sources with a redshift of $z<0.09$ and flux density at 1.4 GHz of
$S_{1400}>10 \,{\rm mJy}$. Additionally, the sample has been extended
to include clusters out to a redshift of $z=0.25$. This radio survey
forms the basis of the present investigation.

As the primary criterion, we have selected sources on the basis of
their morphology.  Extended sources that show clearly the
characteristic `C' shape of WAT sources have been collected to define
a sample of WATs in Abell clusters.  Some radio galaxies which have
been previously classified as WATs by various investigators, but whose
jets do not appear to be significantly bent, are not included in the
present sample. An example of such a source is the radio galaxy
0043+201 in Abell~98; although, for the study of its dynamics it has
been treated as a bent WAT source (Krempec-Krygier \& Krygier 1995),
radio maps do not show the characteristic bent structure (O'Donoghue,
Owen \& Eilek 1990).

Radio galaxies whose jets were found to be smaller than
100$h_{50}^{-1}$~kpc have also been excluded.  In many cases, the optical
light distribution of these galaxies has been found to extend out to
such radii (e.g. Owen \& White 1991, Graham et al.\ 1996).  Since we
are primarily concerned with quantifying the interactions between
radio jets and the ICM, it is important to exclude small sources,
whose jet dynamics are likely to be significantly affected by the
galaxy's own interstellar medium.  The WAT sources PKS 2322$-$123 in
Abell~2597, 2207$-$124 in Abell~2420, 1519+488B in Abell~2064,
1508+059 in Abell~2029, NGC~4874 in Coma, 1142+157 in Abell~1371,
0720+670 in Abell~578 were all excluded on this basis.

The final sample of WATs in Abell clusters is presented in
Table~\ref{tab:WATs}.  The first column lists the names of the radio
source, while the next two [(2) and (3)] give the position of the host
galaxy. In column (4) the Abell cluster that hosts the WAT is listed,
and column (5) gives its redshift.  Additionally, this table gives
details of the available X-ray observations of the field containing
the WAT (see \S 3).

The selection of a radio source and its morphological classification
as a WAT are dependent on the quality and resolution of the available
radio data.  If the resolution of the radio observation is not high
enough to reveal the detailed structure of a WAT, it might be mistaken
for a different class of radio galaxy (such as a narrow-angle tailed
radio galaxy) and not included in the sample.  We have therefore
examined the literature to see if higher quality radio maps exist for
any of the WAT candidates.  Only a few sources were found to have been
observed at higher resolution, most of which are presented by
O'Donoghue et al.\ (1990).  However, since we have restricted the
sample to contain sources that are larger than 100$h_{50}^{-1}$~kpc
and that lie at a redshift of $z<0.25$, there should be very few WATs
that have not been observed with the requisite angular resolution --
at a redshift of $0.25$, a WAT that meets our criteria will have an
extent of $\sim 30$ arcseconds, significantly greater than the spatial
resolution of almost all the radio data.

\begin{figure}
 \begin{center}
 \leavevmode
 \epsfxsize 0.9\hsize
 \epsffile{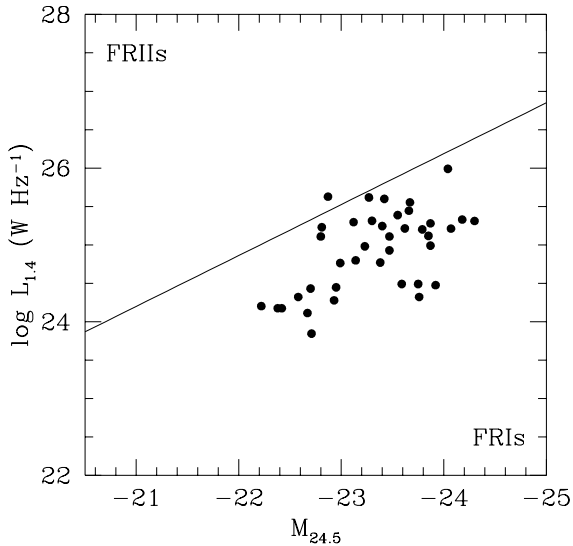}
 \end{center}
\caption{The radio luminosity -- optical magnitude diagram for the
candidate WAT sources. The solid line divides the plane between FR~I
and FR~II radio sources.  This plot is directly comparable to the one
presented by Ledlow \& Owen (1996).}
\label{fig:radio_optical}
\end{figure} 

Confirmation that Table~\ref{tab:WATs} contains a representative
sample of WATs comes from comparing the optical magnitudes of the
galaxies that host them with their total radio power.
Figure~\ref{fig:radio_optical} shows the radio luminosity at 1.4~GHz
versus the absolute magnitude of the galaxies that host the WATs. The
absolute magnitude is taken from Ledlow \& Owen (1995b), and Owen \&
White (1991), and the radio luminosities are calculated from the
fluxes given by Ledlow \& Owen (1995a) using their adopted
cosmology. The plot is directly comparable to the one that was
presented by Ledlow \& Owen (1996), which was constructed using all
the radio galaxies in their sample and others collected from the
literature. The solid line represents the division between galaxies of
Fanaroff-Riley class FR~I and FR~II (Fanaroff \& Riley 1974).
Canonically, WATs are found to be FR~I sources in bright ($-22 >
M_{24.5} > -24.5$) galaxies (O'Donoghue, Eilek \& Owen 1993), and it
is clear that the current sample of sources meets these requirements.

\section{X-ray observations}

We have searched the {\it ROSAT} data archive in order to find which
of the clusters containing WAT sources have been observed in the X-ray
energy band.  The search was constrained to a circle of 30 arcmin
around each radio source.  The {\it ROSAT} observations found and used
in the subsequent analysis are presented in Table~\ref{tab:WATs}
[column (6)]. The sequence numbers designate the detector used for
each particular observation (`wp' or `rp' for PSPC, and `wh' or `rh'
for HRI).  All these datasets, apart from the HRI observation of Abell
160, are publicly available.  Some of the clusters with PSPC
observations have also been observed by the HRI detector, but these
generally-inferior datasets are not reported, or used here. The total
exposure times of the observations are also given in column (7) of
Table~\ref{tab:WATs}.  Notes on the individual sources with X-ray
observations are given in the Appendix.

The clusters were also observed in the {\it ROSAT} All Sky Survey
(RASS) performed with the PSPC. However, as will become clear from the
discussion below, the short exposures in the survey observations mean
that these data are not suitable for the present investigation,

It is clear from Table~\ref{tab:WATs} that less than half of the
clusters in the sample of Abell clusters that contain WATs have been
observed with {\it ROSAT}. In order to increase the coverage of X-ray
observations over the sample, the {\it Einstein} database was also
searched. Only three of the clusters that have not been observed by
{\it ROSAT} have {\it Einstein} observations (A439, A690, A1609).
Unfortunately, in all these cases the exposure times were too short to
reveal any sign of the emission from the ICM of the associated
cluster. Therefore, only the {\it ROSAT} observations could be used in
this analysis.

The sparsity of the X-ray observations reduces the sample of WATs in
Abell clusters.  It also raises the possibility that the clusters for
which X-ray data exist may form a biased subsample of the WAT systems.
In order to check this possibility, we have compared the redshift
distributions, cluster richness distributions, and radio power
distributions of the available subsample and the complete sample.  In
each case, a Kolmogorov-Smirnov test (K-S test; Press et al.\ 1986)
fails to show any evidence that the subsample is in any way biassed.

\begin{table*}
\begin{center}
 \caption{X-ray results}
\label{tab:RESULTS}
\vspace{0.5cm}
\small
 \begin{tabular}{lccccccccc}   \hline \hline
(1) & (2) & (3) & (4) & (5) & (6) & (7) & (8) \\
         & \multicolumn{2}{c}{\underline{Cluster centre (1950)}} &
Positional \\
Cluster  & RA & Dec & error & $d$ & $r_{\rm c}$ & $\theta$&
$\sigma_{|\theta|}$ \\
         & $^{\rm h} \; ^{\rm m}\; ^{\rm s}$ & $\degr \; \arcmin \; \arcsec$ & (arcsec/kpc)  & (arcmin/$h_{50}^{-1}$kpc) &
($h_{50}^{-1}$kpc) & (degrees) & (degrees)\\ 
\hline
 
A160 & 01 10 20.2 & +15 15 04.6 & 25/32 & 1.49/115.3 & -- & --111.1 &
15.5 \\

A194 & 01 23 10.5 & --01 36 34.6 & 30/16 & 4.27/133.2 & 296$\pm$39 &
2.3 & 6.8 \\

A400 & 02 54 57.7 & +05 47 40.1 & 30/21 & 2.25/93.1 & 236$\pm$5 &
--0.2 & 12.7 \\

A562 & 06 47 57.9 & +69 24 08.1 & 5/16 & 0.68/130.6 & 144$\pm$13$^{g}$
& 39.1 & 7.0 \\

A569 & -- & -- & -- & -- & -- & \\
 
A623 & 08 03 05.1 & --00 48 50.0 & 30/76 & 0.88/133.6 &
162$\pm$57$^{g}$ & 96.7 & 29.6 \\

A1314 & 11 32 04.5 & +49 23 03.4 & 20/20 & 8.33/489.8 & 322$\pm$40 &
35 & 2.3 \\

A1446 & 11 59 27.3 & +58 19 25.2 & 5/15 & 0.70/126.4 &
350$\pm$17$^{g}$ & --13.5 & 6.8 \\

A1552 & 12 27 39.5 & +12 01 12.0 & -- & 6.16/905.5 & -- & --14.5 & -- \\

A1569 & 12 33 50.5 & +16 49 31.7 & 40/20 & 1.34/183.3 &
268$\pm$162$^{g}$ & --3.3 & 6.2 \\

A1763 & 13 33 06.5 & +41 15 09.3 & -- & 0.62/246.6 & 428$\pm$10 & 176.1
& -- \\

A1890 & 14 15 05.4 & +08 26 39.8 & 50/83 & 0.70/69.7 & 301$\pm$130 & 110.8 &
50.0 \\

A1940 & 14 33 43.4 & +55 21 00.9 & 15/61 & 1.64/399.5 &
304$\pm$145$^{g}$ & 143.5 & 8.7 \\

A2214 & 16 36 13.3 & +37 58 54.3 & -- & 0.48/134.8 & 263$\pm$32$^{g}$
& 36.1 & -- \\
      & 16 35 47.2 & +37 59 13.9 & -- & 5.63/1580.9 & & 33.9 & -- \\

A2220 & 16 38 44.6 & +53 52 54.7 & -- & 2.99/577.7 & 328$\pm$43 &
--4.2 & -- \\

A2304 & 18 20 43.4 & +68 56 03.5 & -- & 3.84/589.8 &289$\pm$32$^{g}$ &
20.2 & -- \\

A2306 & 18 26 26.8 & +74 42 48.1 & 20/74 & 0.74/164.3 &
133$\pm$16$^{g}$ & --8.0 & 24.2 \\

A2462 & 22 36 29.7 & --17 36 11.9 & 10/20 & 0.15/18.3 &
332$\pm$22$^{g}$ & 143.6 & 47.5 \\

A2634 & 23 35 53.6 & +26 43 56.8 & 15/14 & 1.78/99.3 & 310$\pm$16 &
155.7 & 8.0 \\

A2637 & -- & -- & -- & -- & -- & \\
\hline \\
\multicolumn{4}{l}{NOTE: $^{g}$ core radii calculated by G\'{o}mez et
al.\ (1997b)} \\ 
 \end{tabular}
\end{center}
\end{table*}

\section{Data analysis}

\subsection{X-ray centres}

In order to quantify how far a WAT is offset from its host cluster's
centre, we need an objective definition of the centroid of the X-ray
emission.  As mentioned in the Introduction, in a merger one would
expect the details of the X-ray emission to be rather complicated, so
we need a measure of the cluster centroid that is insensitive to this
complexity.  We discuss the possible impact of such complexity on the
analysis later in this section.  Given the short exposures in some of
the X-ray observations, we also need to be sure that the method used
for calculating the cluster centre provides a robust estimate, even
when the quality of the data is rather low.  Finally, it is 
important that we adopt an objective process: if we were to examine
the radio maps before estimating the X-ray centroid, there would be a
danger that we might bias our results by, for example, picking out the
X-ray peak that happens to lie closest to the WAT.  Such a
reinforcement of any pre-existing prejudice must be avoided, so, as
far as possible, we have carried out the X-ray centroid calculations
without prior reference to the WAT location, and using an algorithm
that requires as little human intervention as possible.  In this
section, we discuss the adopted procedure, and the resulting
determinations of the cluster center location.

Having first removed any bright point sources from the X-ray data by
interpolation, the task we are faced with is quantifying the
distribution of diffuse gas so as to define its centroid.  After some
experimentation, it became apparent that the most robust way of making
such a quantification was also pretty much the simplest.  The emission
from the central region of the cluster was projected on to the $x$- and
$y$-axes of the image, and the counts on each projected axis were
fitted by a Gaussian function.  In each fit, the amplitude, width, and
centre of the Gaussian were left as free parameters.  The best fit
model provides, along with the best-fit values of the other free
parameters, the centre of the Gaussians in each axis; these numbers
provide a surprisingly robust measure for the location of the centre
of each cluster.  

The cluster X-ray centres calculated in this way are presented in
Table~\ref{tab:RESULTS} [columns (2), (3)].  This procedure also
provides the errors of the determined cluster centre, which are given
in Table~\ref{tab:RESULTS} [column (4)]. The results and their
position on the X-ray images were also examined by eye, to make sure
that the adopted centre had not been unduly influenced by any
residuals from the subtraction of the point sources.  It is important
to stress that we make no claim that these simple Gaussian fits
describe the actual distribution of X-ray emitting gas; rather, the
fitted parameters simply provide a robust and objective estimator for
the location of the X-ray centroid.  We visually inspected each
derived centroid to check that the process really does provide a
credible measure of the centroid; in almost all cases, the derived
location was found to agree well with a ``$\chi$-by-eye'' fit, but
without the danger of statistical biasses inherent in the latter
subjective process.

In a few cases, as described in more detail in the Appendix, such a
procedure proved unfeasible, mainly because the cluster is clearly
bimodal, so the whole concept of `a cluster centre' is flawed, and a
single Gaussian cannot be fitted unambiguously to the data.  In such
cases, the brightest peak of the X-ray emission was taken to indicate
the centre of the cluster.

The centres of some of the clusters that host WATs have previously
been determined by Briel \& Henry (1993), Pierre et al.\ (1994), and
Ebeling et al.\ (1996) using the RASS. Comparison of their results to
the positions reported in Table~\ref{tab:RESULTS} supports the concern
mentioned in \S3 that the short exposures in the RASS could lead to
inaccurate determinations of the cluster centres. In several cases the
centres that are given by these investigations coincide with the
optical galaxy that hosts the radio source.  Since radio galaxies are
often also strong X-ray emitters, it is very likely that in these cases
the analysis based on RASS data picked up the location of the emission
coming from the WAT rather than the centre of the ICM distribution.

The X-ray light distributions of the clusters that host WATs generally
appear somewhat irregular, and elongated (a point noted by G\'{o}mez
et al.\ 1997b).  These observations are clearly not consistent with the
simple picture of the ICM forming a spherically-symmetric structure,
and some care must be taken in interpreting the data.  However, the
procedure adopted here for determining the centre of the X-ray
distribution provides a robust estimator for the centroid of the
emission even for elongated and irregular clusters, since it does not
attempt to follow any small scale irregularities in the data.

One interpretation of the irregularities in the X-ray emission is that
it may well reflect the recent merger between clusters that might also
be responsible for producing the WAT.  This interpretation raises a
further concern: the violent hydrodynamical processes that occur in
the merger between two clusters' ICMs mean that the centroid of the
X-ray emission need not coincide with the merger remnant's centre of
mass, so that it may not provide a good fiducial measure of the
cluster centre.  Hydrodynamic simulations of merging clusters of
galaxies (Roettiger, Stone \& Mushotzky 1997, and references there-in)
confirm that such shifts do occur.  However, even for the most extreme
case of a collision between two equal-mass clusters, the shift is much
less than the cluster core radius.  As we shall see below, such shifts
are small compared to most other distances in this analysis, so the
X-ray centroid definitely provides an adequate measure of a cluster's
centre for the current study.

\subsection{Size of the cluster}

The WAT radio sources presented here are located in a variety of
environments, from poor (richness class $R = 0$) clusters up to
relatively rich ($R = 3$) systems.  The extent of the clusters might
be expected to vary accordingly.  Therefore, if we are to compare
results for different WATs, it may be useful to scale distances by the
sizes of their host clusters.  In this study, where we are interested
in the impact of the ICM on the jets, a sensible scale-length is
provided by the core radius of the distribution of the ICM.

Core radii of some of the clusters that host the WATs have been
previously measured by G\'{o}mez et al.\ (1997b), using the {\it ROSAT}
PSPC observations. They fitted the surface brightness distribution of
each cluster by the traditional $\beta$-model, leaving the central
surface brightness, the core radius, and the $\beta$ parameter to be
determined by the fit.  Their calculated values for the core radii,
converted from their choice of cosmology to the one adopted here, are
given in Table~\ref{tab:RESULTS}.

For the remainder of the clusters, whose X-ray observations have not
been previously analysed, we have carried out a similar procedure.
Counts were integrated in concentric annuli, centered on the cluster
centre as found in \S4.1.  The width of each annulus was different
for each cluster, depending on the number of photons detected.  All
the point sources lying on the image of the clusters were masked out.
The radial profile was then fitted by the $\beta$-model, with the
background left as free parameter to be determined by the fit. The
limited integration time for most of these observations prevented us
from satisfactorily fitting for both $\beta$ and $r_c$, so we fixed
$\beta = 0.65$, which is an average value for this parameter found
from the study of other similar clusters (Jones \& Forman 1984).  The
resulting values of the core radii are given in column (6) of
Table~\ref{tab:RESULTS}.

\section{Results}

\subsection{The spatial distribution of WATs in clusters}

\begin{figure}
\begin{center}
 \leavevmode
 \epsfxsize 1.0\hsize 
 \epsffile{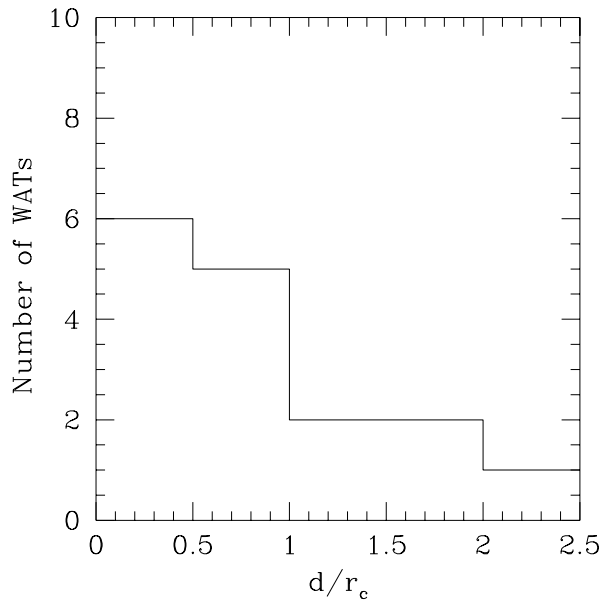}
\end{center}
\caption{The distribution of distances of WAT sources from their
cluster centres, measured in units of the clusters' core radii.} 
\label{fig:distance}
\end{figure}

Having measured the X-ray location of a cluster's centre, we can now
quantify the offset between the location of a WAT and its cluster
X-ray centroid.  The distribution of observed distances scaled with
the core radius of each cluster is presented in
Fig.~\ref{fig:distance}.  It is apparent that, although WATs are found
preferentially toward the centres of clusters, they are spread over a
wide range of distances from the cluster centre.  In fact, this figure
does not include the most extreme case of Abell~2214, where the WAT
lies at $\sim 6$ core radii.  It should also be borne in mind that
these offsets are even more significant once projection effects are
taken into account, since the observed projected radius of a WAT only
places a lower limit on its true distance from the cluster centre.
Thus, we have found strong confirmation that WATs are not all located
close to the centres of their host clusters.


\subsection{Orientation of WATs}

Since we have shown that the galaxies which host WATs often lie far
from the X-ray centroids of the surrounding clusters, it is at least
plausible that the radio jets in WATs are bent by ram pressure as
their host galaxies move relative to the ICM.  In this section, we
attempt to further investigate this scenario by using the direction in
which the jets are bent to determine the WAT galaxies' orbits.  This
approach has previously been taken -- and its limitations discussed --
by O'Dea et al.\ (1987); they explored the dynamics of clusters using
the morphology of the more dramatically-bent radio galaxies known as
narrow-angle tailed sources.

The parameter of interest for such an investigation is the angle
between the line connecting the cluster centre to the radio galaxy,
and the line that bisects the angle between the two radio lobes.  We
define this angle, $\theta$, to be measured counterclockwise from the
radius vector that connects the cluster centre to the optical
galaxy. If the radio jets are bent by ram pressure, then such a
definition assigns $\theta = 0$ for a galaxy travelling directly
toward the cluster centre on the plane of the sky, and $\theta =
180\,{\rm deg}$ for one travelling radially away from the cluster
centre.  The values for $\theta$ obtained from the measured X-ray
cluster centroids, the locations of the WATs, and their observed radio
morphologies are presented in Table~\ref{tab:RESULTS}.

For each system, the error in $\theta$ mainly depends on the accuracy
of the position of the cluster centre; the position of the optical
galaxy has been measured with an accuracy of 1 arcsec (Ledlow \& Owen
1995b). For the sources for which a measurement of the error for the
position of the cluster centre exists, the error in $\theta$ is
calculated and presented in Table~\ref{tab:RESULTS}.  Generally, the
largest inaccuracies of $\theta$ correspond to sources which lie very
close to the cluster centre, since a small change in the position of
such a source corresponds to a large change in angle.

\begin{figure}
\begin{center}
 \leavevmode
 \epsfxsize 1.0\hsize 
 \epsffile{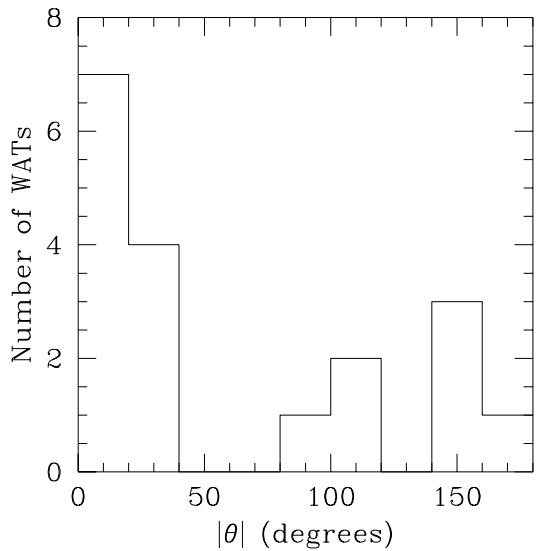}
\end{center} 
\caption{The distribution of the observed angles for all the 18
WATs.}             
\label{fig:angle}
\end{figure} 

Figure~\ref{fig:angle} shows the distribution of $\theta$ for the WAT
sources in this sample.  Since the geometry is symmetric about $\theta
= 0$, we have plotted $|\theta|$, with the angle defined in the range
$-180 < \theta < 180$ degrees.

It is apparent from Figure~\ref{fig:angle} that there seems to be a 
concentration of WAT sources with $\theta \sim 0$ degrees.  We
therefore now investigate whether this concentration might arise from
a statistical fluctuation in the small number of observations, and, if
not, what distribution of orbits might give rise to such a
distribution. 

If the orbits of WATs were entirely random and isotropic, one would
expect to observe all angles $\theta$ with equal probability.  When
the distribution of $|\theta|$ in Figure~\ref{fig:angle} is compared
with such a uniform distribution using the K-S test, the hypothesis
that they have the same distribution can be rejected at the 99.9\%
confidence level.  Thus, the spike at $|\theta| \sim 0$ is
statistically significant, and allows us to confidently rule out the
possibility that WATs follow purely random orbits.

We can exclude the possibility of circular orbits with even greater
confidence.  The distribution of $\theta$ that one would expect for
circular orbits is somewhat more complicated due to projection
effects.  O'Dea et al.\ (1987) have quantified this distribution, and
they found that the expected number of galaxies has a minimum at
$|\theta| = 0$ degrees, and rises to a sharp maximum at $|\theta| = 90$
degrees, before dropping back to a minimum at $|\theta| = 180$
degrees.  Not surprisingly, this distribution is inconsistent with the
distribution in Figure~\ref{fig:angle} which peaks at $|\theta| = 0$
and $180$ degrees; a K-S rejects the possibility that the orbits of
WATs are circular at the 99.99\% confidence level.

\begin{figure}
\begin{center}
 \leavevmode
 \epsfxsize 1.0\hsize 
 \epsffile{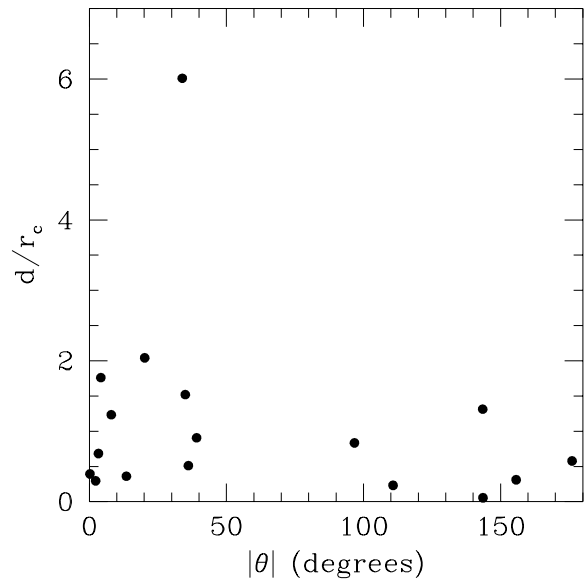}
\end{center} 
\caption{The distance of each WAT source from the cluster centre
versus the angle of its motion as indicated by its jets. The WATs in
Abell~160 and Abell~1552 are not included because the existing X-ray
data do not allow a reliable determination of $r_c$ for these
systems.}
\label{fig:d_angle}
\end{figure} 

We can therefore conclude that the orbits of WATs are predominantly
radial.  If the orbits were purely radial, then the distribution of
$\theta$ should consist of delta-function spikes at $|\theta| = 0$ and
180 degrees.  Since there are observational uncertainties in the
values of $\theta$, we would expect the distribution to be somewhat
broadened.  The presence of a few WATs with intermediate values of
$\theta$ suggests that the orbits cannot be completely radial.
However, it is notable from Figure~\ref{fig:d_angle}, which plots the
distance of the WAT from the cluster X-ray centre as a function of
$|\theta|$, that the WATs at intermediate angles tend to lie at small
cluster radii, where the uncertainties in the values of $\theta$ are
greatest.

\subsection{In-coming versus out-going galaxies}

A tail angle of $0 < |\theta| < 45$ degrees indicates that the WAT is
moving towards the centre of the cluster (in-coming), while an angle
of $135 < |\theta| < 180$ degrees implies an out-going WAT.
Figure~\ref{fig:angle} gives the distinct impression that there are
more WATs travelling toward the centres of clusters than there are
outward-bound systems.  In fact, there are $n_{\rm in} = 11$ in-coming
WATs and only $n_{\rm out} = 4$ out-going systems, and a binomial
distribution with $p = 0.5$ will produce such an imbalance only $\sim
6\%$ of the time.  Thus, we can conclude that the difference between
the observed number of in-coming and out-going WATs is significant at
a level of more than 90\%.

\begin{figure}
\begin{center}
 \leavevmode
 \epsfxsize 1.0\hsize 
 \epsffile{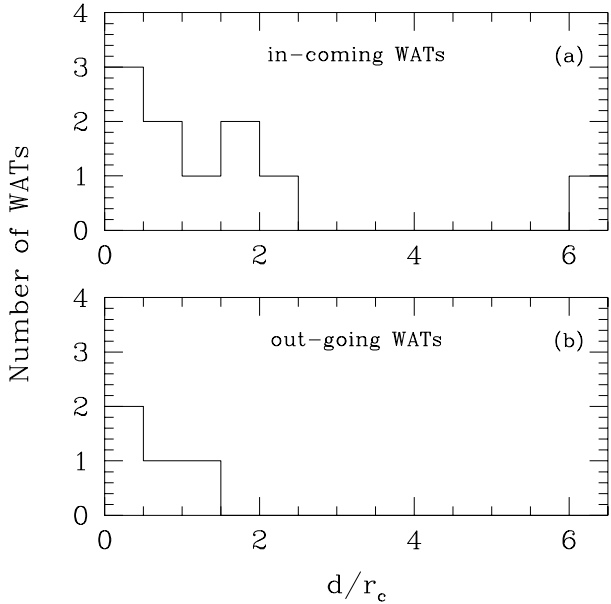}
\end{center}
\caption{The distribution of distances from the cluster centre: (a)
for in-coming sources; and (b) for out-going sources.}
\label{fig:inout} 
\end{figure} 

A further impression one gains from Figure~\ref{fig:d_angle} is that
out-going WATs are found closer to the cluster centre than the
in-coming ones.  Figures~\ref{fig:inout}~(a) and (b) show the radial
distribution of in-coming and out-going WATs respectively. The 10
in-coming WATs with measured values of $r_c$ lie at a mean radius of
$(1.5 \pm 0.5)r_c$, while the 4 out-going have a corresponding mean
radius of $(0.6 \pm 0.3)r_c$.  Applying Student's $t$-test to these
data, we find that the two means are significantly different at the
95\% confidence level.  However, it should be pointed out that this
result is not very robust: if we exclude the single in-coming WAT with
$d = 6 r_c$ (in Abell~2214), then the difference ceases to be
statistically significant.

\section{Summary and Discussion}

We have defined a sample of WAT radio sources in Abell clusters in
order to investigate the hypothesis that these sources are bent by ram
pressure induced when their host galaxies are kicked into motion by a
cluster merger.  Archival {\it ROSAT} observations have been used to
define more accurately and objectively the X-ray centres and sizes of
the clusters that host the WATs.

The basic findings of this analysis are as follows:
\begin{enumerate}

\item WATs are not generally located at the centres of their host
clusters as defined by their X-ray emission.  They are found over a
range of distances from the cluster centre, out to several core
radii. 

\item If their bent shape can be attributed to ram pressure, then WATs
are found to lie preferentially on radial orbits.

\item There are more WATs travelling toward the centres of their host
clusters than there are systems moving away from the centres.

\item There are indications that WATs travelling toward the centres of
clusters lie at larger radii on average than those travelling outward.

\end{enumerate}

These findings are exactly what one would expect if WATs are created
by mergers between clusters.  Specifically:
\begin{enumerate}

\item When two clusters merge, the D/cD radio galaxy that would
initially have lain at the centre of one of the merging systems will no
longer be at the centre of mass.  Whatever the localized impact of the
collision on the X-ray emission, one would not expect to find the
WAT near the new X-ray centroid.

\item In such a merging system, the radio galaxy will continue to move
in the direction that its host cluster was travelling in prior to the
merger.  Such mergers will largely arise from the gravitational
attraction between the two pre-existing clusters, resulting in a
head-on collision.  The radio galaxy will therefore travel along the
line joining the centres of the two merging systems.  In terms of the
merged system, it will therefore initially be moving toward the new
cluster centroid, on a radially infalling orbit.  The X-ray emission
will be somewhat disturbed by the merger process, but, as we have
discussed in \S4.1, the centroid of the emission will be shifted by
very much less than the new cluster's core radius.  Since the typical
distances between the cluster centroid and the infalling WAT are at
least comparable to the core radius (see Fig.~\ref{fig:distance}),
this shift is pretty much negligible, and we would expect to see the
galaxy's radio jets bent away from the X-ray centroid.  Perhaps the
best argument for the validity of the way in which the centroid has
been estimated comes from Fig.~\ref{fig:angle}: if the adopted cluster
centres were in error to the point that they had no physical meaning,
then comparison with the independently-derived WAT morphologies could
never lead to the correlation shown in this figure.

\item Near the new cluster centre, the gas density will be high, and
there will be a complex structure of shock-heated gas and turbulence
due to the collision.  Such an environment will prove very hostile for the
relatively fragile radio jets, either destroying the jets entirely or
disrupting them to a point where the system is no longer identified as
a symmetric WAT.  The destruction of a significant fraction of WATs as they
pass near the cluster centre explains the imbalance between the
numbers of in-coming and out-going systems.

\item The D/cD galaxy that host a typical WAT is so massive that dynamical
friction will play an important role, even on the galaxy's first
passage through the merged cluster's centre.  For example, in their
numerical simulations of cluster formation through heirarchical
mergers, Frenk et al.\ (1996) found that massive infalling galaxies
with typical initial velocities in excess of $900\,{\rm km}\,{\rm
s}^{-1}$ were slowed to a mere $\sim 200\,{\rm km}\,{\rm s}^{-1}$ on
their first passage through the cluster core [see Frenk et al.\ (1996)
Figure~9; for example, their galaxy 3].  Such decelerated infalling
galaxies will not travel back out to anything like the radius from
which they initially arrived [see Frenk et al.\ (1996) Figure~8],
explaining the difference between the mean radii of the in-coming and
out-going WATs.  It is also notable that velocities of a
few hundred kilometres per second in the typical gas density near a
cluster core are exactly what one needs to bend a radio source into a
WAT [see, for example, Sakelliou et al.\ (1996)].  

\end{enumerate}

This analysis therefore provides very strong support for the
hypothesis that the morphology of a wide angle tail radio source
results from ram pressure against the surrounding intracluster medium,
and that the impetus for the radio galaxy's motion has come from a
recent cluster merger.

\section*{ACKNOWLEDGMENTS} 
 
This research has made use of the NASA/IPAC Extragalactic Database
(NED) which is operated by the Jet Propulsion Laboratory, California
Institute of Technology, under contract with the National Aeronautics
and Space Administration. Much of the analysis was performed using
{\sc iraf}, which is distributed by NOAO, using computing resources
provided by STARLINK.  MRM has been supported by a PPARC Advanced Fellowship
(B/94/AF/1840).

\appendix

\section{Notes on individual sources}

{\it Abell~160:} The X-ray structure of this cluster is
irregular, showing distinctive condensations. From early {\it
Einstein} observations these clumps of X-ray emission have been
identified with emission from the cluster's galaxies. The {\it ROSAT}
HRI image reveals the same situation. The definition of the X-ray
centre is rather difficult, even after the removal of the bright point
sources. The present data do not permit the calculation of the core
radius, since the $\beta$-model does not provide a good fit to the surface
brightness distribution of the ICM.  Additionally, the jets of the
radio galaxy are not severely bent, a fact that indicates the lack of
a very dense ICM.

{\it Abell~400:} This cluster hosts the extraordinary
radio source 3C~75, which consists of a dumbbell pair of radio
galaxies. The jets of both radio galaxies are bent in the same
direction, suggesting that they are both bent by the same cause. The
interpretation of Balcells et al.\ (1995) that this source is the
result of a merger of two different clusters, where each cluster
hosted one radio galaxy does not look plausible. In such a scenario
it is difficult to explain the bending of the jets in the same
direction. This source is treated as one radio galaxy in the present
survey. 

{\it Abell~569:} The PSPC image is dominated by the
emission from galaxies and point sources. The distribution of the
cluster's ICM is not clearly revealed.

{\it Abell~690:} The cluster and radio galaxy lie underneath the rib
suport structure of the PSPC detector.

{\it Abell~1552:} This cluster lies behind the Virgo
cluster. Therefore, its X-ray emission is contaminated by the emission
from the ICM of the Virgo cluster. The cluster centre reported here
coincides with the position of the brightest galaxy of the
cluster. This galaxy is also a radio galaxy, and the radio maps (Owen
\& Ledlow 1997) show that its jets are not distorted, which implies
that the galaxy is not in motion relative to the ICM. An attempt to
fit only the southern part of the cluster with the $\beta$-model
yielded inconsistent results, and therefore, a measurement of the core
radius cannot be provided.

{\it Abell~1763:} The redshift of this cluster that has
been extensively used (0.187; Struble \& Rood 1991) was originally
calculated using the redshift of only one galaxy (Noonan
1981). Recently, Owen, White \& Thronson (1988) and Owen, Ledlow \&
Keel (1995) have measured the redshift of the galaxy that hosts the
WAT and find it to be 0.2278, which is the value that is used here.
The cluster centre that is reported here is the peak of the X-ray
emission.

{\it Abell~1890:} The X-ray emission from this relatively
poor cluster appears to be clumpy in this short HRI
observation. 

{\it Abell~2214:} This cluster is clearly bimodal. Thus, a
calculation of the cluster centre would be misleading. Both peaks of
the X-ray emission are used for the definition of the cluster
centre. However, the choice of a cluster centre does not influence
much the value of the angle $\theta$, since the position of the WAT is
nearly aligned with both peaks.

{\it Abell~2220:} This cluster appears bimodal in the {\it
ROSAT} image.  Apart from a peak of X-ray emission that coincides with
the galaxy that hosts the WAT, there are two more aligned peaks of X-ray
emission. The position of the middle peak coincides with a big
(non-active) galaxy which belongs to Abell~2220, while the eastern
peak does not have any pronounced optical counterpart in the Palomar
plates.  For this reason, the middle peak is adopted as the cluster
centre. In any case, this choice for the cluster centre does not influence
the measurement of the angle $\theta$, since both peaks are aligned
with the position of the galaxy that hosts the WAT. Additionally, the
present selection puts the WAT nearer to the cluster centre.

{\it Abell~2304:} The calculated cluster centre coincides
with the peak of the X-ray emission.

{\it Abell~2306:} The X-ray image is clumpy. 

{\it Abell~2634:} A detailed study of the hot gas context
of this cluster, and investigation of the mechanisms responsible for
the features of the radio jets of 3C465 can be found in Sakelliou \&
Merrifield (1998a), Schindler \& Prieto (1997), Sakelliou \&
Merrifield (1998b). 

{\it Abell~2637:} There is no sign of X-ray emission
from the cluster in the available X-ray image.

\end{document}